# Nanosystem Self-Assembly Pathways Discovered via All-Atom Multiscale Analysis

*Stephen D. Pankavich**

Department of Mathematics

United States Naval Academy

Annapolis, MD 21402

*Peter J. Ortoleva*

Center for Cell and Virus Theory

Department of Chemistry

Indiana University

Bloomington, IN 47405

*Corresponding Author: pankavic@usna.edu

ABSTRACT

We consider the self-assembly of composite structures from a group of nanocomponents, each consisting of particles within an *N*-atom system. Self-assembly pathways and rates for nanocomposites are derived via a multiscale analysis of the classical Liouville equation. From a reduced statistical framework, rigorous stochastic equations for population levels of beginning, intermediate, and final aggregates are also derived. It is shown that the definition of an assembly type is a self-consistency criterion that must strike a balance between precision and the need for population levels to be slowly varying relative to the timescale of atomic motion. The deductive multiscale approach is complemented by a qualitative notion of multicomponent association and the ensemble of exact atomic-level configurations consistent with them. In processes such as viral self-assembly from proteins and RNA or DNA, there are many possible intermediates so that it is usually difficult to predict the most efficient assembly pathway. However, in the current study rates of assembly of each possible intermediate can be predicted. This avoids the need, as in a phenomenological approach, for recalibration with each new application. The method accounts for the feedback across scales in space and time that is fundamental to nanosystem self-assembly. The theory has applications to bionanostructures, geomaterials, engineered composites, and nanocapsule therapeutic delivery systems.

# Introduction

The self-assembly of composite structures from nanocomponents is central to many natural and engineered processes. Viruses self-assemble from macromolecules and their associated complexes[1]. Smart materials, molecular machines, ribosomes, and molecular electronics are other examples of self-assembling nanostructured materials[2-4], as are geological materials such as opal. Hence, understanding self-assembly pathways and the factors that control them is of great interest in the pure and applied sciences. In the present study, a theory of self-assembly is derived from the laws of molecular physics. Our motivation is to both attain a deeper understanding of the pathways involved and to arrive at a calibration-free theory. In short, we seek an understanding of the interplay of processes across multiple scales in space and time that underlies self-assembly.

Modeling the self-assembly of composites from nanocomponents presents numerous conceptual and practical challenges. Each nanocomponent consists of many highly fluctuating atoms, and hence interactions among components are not well defined. This is in contrast to interatomic forces that have been formulated and well calibrated with experimental data and quantum computations that provide unique values of atomic forces given a corresponding configuration. Nanocomponents have a highly fluctuating internal state, which underlies their entropics and average energetics. Furthermore, the internal structure and fluctuations of one nanocomponent can be strongly affected by the presence of others. Thus, fluctuations and frictional/dissipative effects play a key role in the self-assembly and stability of nanocomposites. Similarly, the position, orientation, and nanoscale internal structure of the components affect assembly in a complex fashion as they modify statistical properties of fluctuations within the composite and its microenvironment. Conditions in the microenvironment (i.e., pH, salinity, selected ion concentrations, and temperature) alter the internal state and the evolving nature of interactions between nanocomponents. From this, we conclude that a predictive calibration-free theory of self-assembly should be based on an underlying all-atom description for which all forces are well-understood.

Kinetic theories of self-assembly have been presented for many of the aforementioned systems[5-9]. As for chemically reacting systems, phenomenological kinetic theories require extensive experimental data and prior knowledge of steps along the self-assembly pathway. This severely limits the predictive power of such approaches. The well-known variety of assembly pathways (e.g. $A+B \rightarrow AB, AB+C \rightarrow ABC$ versus the alternative permutations of $A, B$ and $C$) illustrate the factorial multiplicity that could emerge when assembly involves many components. While phenomenological kinetic studies have furthered the understanding of self-assembly in well-characterized systems, the detailed kinetics of the processes in most systems remains unclear, especially considering the highly dissipative and stochastic nature of the Brownian dynamics of the nanocomponents involved. When so much experimental data is needed, the question arises as to the meaning of "prediction". To arrive at a complete theory, all intermediate structures along the assembly pathway must be identified and rate data for individual assembly steps must be obtained. Given this we conclude that a theory of self-assembly should have the following characteristics:

1. the underlying description is all-atom in detail to arrive at a calibration-free model
2. the internal atomic structure of the nanocomponents adjusts dynamically as the many-component system changes so that inter-component forces must be co-evolved with the many-component state
3. the population levels of initial, intermediate, and final assemblies are co-evolved with the probability density of atomistic configurations to capture the multiscale character of nanocomposite self-assembly
4. some intermediate assemblies may be rare; thus population fluctuations can be significant, requiring that the theory must be stochastic in nature[10] and allow for multiple outcomes or particular assembly pathways beginning with the same initial population levels of individual nanocomponents
5. criteria are provided to determine the completeness of the description and impose restrictions on characteristics of the nanocomponents involved for the self-consistency of the theory.

The objective of this study, then, is to develop a theory of self-assembly with these characteristics.

Since self-assembly is ultimately directed by the laws of molecular physics, an all-atom description is the natural starting point for deriving the kinetic theory. Given the broad range of characteristic times in self-assembly (e.g., $10^{-14}$ seconds for atomic collisions/vibrations versus $10^{-3}$ seconds or longer to form complete assemblies), we adopt a multiscale formalism. In this framework, order parameters (i.e., population levels of long-lived assemblies) are defined in terms of the $6N$ atomic positions and momenta, $\Gamma$, and describe overall properties of the system. Further, they are shown via Newton's equations to evolve on a timescale much greater than that of individual atomic vibration. Since the order parameters depend on $\Gamma$, they do not evolve independently of atomic variables. Hence, the order parameter evolution alone does not provide a complete theory of population dynamics.

One technique that has been proposed to address this incompleteness dilemma is the projection operator method of Mori[11] and Zwanzig[12]. However, this method presents practical and conceptual difficulties. Projection operators involve the integration of a subset of atomic variables from the $N$-atom probability density $\rho$, and the method arrives at a formal equation satisfied by a reduced density for the variables of interest (population levels here). This formal equation involves expressions (memory kernels) for which the time-dependence is generated by an evolution operator $e^{(\mathbb{I}-P)\mathcal{L}t}$ where $P$ is a projection operator, $\mathcal{L}$ is the Liouville operator, and $t$ is time. This evolution operator acts on functions of $\Gamma$. Unfortunately, evolution of this type is difficult to simulate as no software like molecular dynamics exists for carrying out the projection operator-mediated evolution. In fact, if the associated memory kernels decay on a timescale that is longer than that of atomistic changes, then they are at least as intensive to compute as solving the original Liouville equation itself, or more realistically using MD for the entire computation. Progress can be made when there is a timescale separation, but this raises the question as to why the projection operators should have been used in the first place. For instance, why not use a multiscale perturbation scheme directly to utilize the timescale separation?

There is an extensive literature on the use of multiscale techniques to analyze the Liouville equation[13-23]. The present work is an extension of that approach, notably the all-atom multiscale (AMA) method[24-33]. Frictional effects in the dynamics of many nanoparticle systems undergoing aggregation have been considered using a multiscale approach[34-36]. Such effects are also included in the work presented here but via a distinct formulation involving population levels as collective variables.

The AMA method adopts a direct perturbative strategy[24-33]. In AMA the Liouville equation, originally cast in 6$N$ dimensions, is mapped to a higher-dimensional product space using a multiscale ansatz on the $N$-atom probability density. Thus, $\rho$ is hypothesized to have the dependence $\rho(\Gamma,\Phi,t)$ for a set of $N_a$ order parameters $\Phi$, and hence is a function in $(6N+N_a)$-dimensional state space. With the chain rule and the $\Gamma$-dependence of $\Phi$, one obtains a reformulated Liouville equation for $\rho$ as an explicit function of both $\Gamma$ and $\Phi$. This unfolded Liouville equation is solved as a power series in a timescale ratio $\varepsilon$. The parameter $\varepsilon$ is a ratio of characteristic lengths, times, or masses and naturally emerges due to the slow variations of the order parameters relative to the timescale of atomic collisions. Upon defining the reduced probability density $\Psi(\underline{\Phi},t) = \int d\Gamma^* \delta(\underline{\Phi}-\underline{\Phi}^*)\rho(\Gamma^*,\underline{\Phi}^*,t)$ an exact conservation law results from the structure of $\rho$. Then, using the multiscale solution of the Liouville equation, a closed equation of Fokker-Planck or Smoluchowski type for the state of the $N_a$ order parameters $\Phi$ is obtained, and the need for projection operators is avoided.

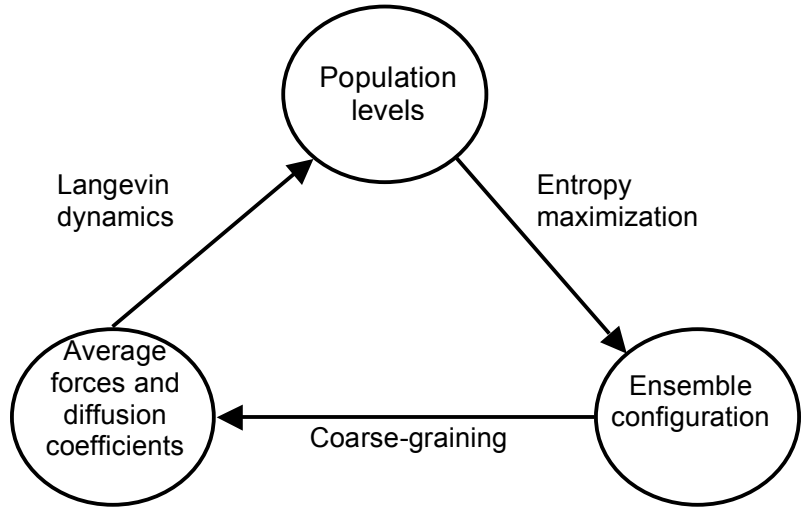

**Fig. 1:** Population levels characterize the overall organization of a self-assembling system, which affects the probability of the atomistic configurations. This mediates thermal-average forces driving population dynamics, and diffusion factors mediating the rate of change and coupling among the population levels. This interscale feedback loop is central to understanding nanosystem self-assembly. The proposed multiscale approach captures this feedback by co-evolving population levels with the thermal-average forces and diffusion factors, thereby providing the essence of an algorithm for the numerical simulation of self-assembly.

Coarse-grained models have also been proposed but are built entirely upon equations for a reduced set of variables (i.e., $< 6N$). It is incorrect to term these approaches "multiscale" unless they track both the atomistic and the course-grained descriptive variables. Thus, we term such approaches "decoupled coarse-graining" when they do not co-evolve dynamics at all scales. The objective of decoupled approaches is to reduce the number of variables (e.g., to describe clusters of atoms) using heuristic arguments and calibration with experimental data to fit the number of degrees of freedom and computational time. Unfortunately, such methods do not account for the key interscale feedback suggested in Fig. 1. In reality, the order parameters mediate the ensemble of atomic-scale configurations while the latter determine the free-energy gradients and diffusivities that drive order parameter dynamics. In the decoupled approach, effective interaction between clusters of atoms[37-38] or between an atomically described zone and a continuum[39-40] are calibrated prior to simulation. Thus, the feedback of Fig. 1 is ignored. In contrast, our multiscale approach co-evolves the probability density of atomistic configurations along with order parameters. Decoupled approaches assume the lumped elements evolve via a Langevin equation. However, elements are usually too small to justify the timescale separation required for this approach. AMA displays that this requires the ratio of the mass of a typical atom to that of a lumped element is small. But if the elements are large, it is necessary to account for the internal atomistic fluctuations and associated dissipative nature of collisions of larger elements. Thus, in the presence of a distinct characteristic scaling AMA provides a precise multiscale description of the system.

A self-consistency check is also provided by the all-atom analysis. In the stochastic equation of order parameter dynamics, diffusion factors appear. These factors are related to correlation functions of order parameter velocities. The correlation functions can be readily computed via MD when the correlation time is short relative to the characteristic time of order parameter evolution. If key order parameters are missing from the analysis, these correlation functions can have long-time tails. In this case, unacceptably large or even divergent diffusivities may arise. This difficulty can then be remedied by including more order parameters in the description.

A challenge for the theory of self-assembly is that there is no clear notion of species (i.e., assembly types). Each nanocomponent in an assembly could be in any of a number of nanometer and atomic scale configurations, and similarly for the architecture of an assembly of these components. Thus, an assembly type (species) is best conceived of as an association of a given set of atoms with a broad spectrum of possible configurations. However, as the definition of the common structural feature shared by the ensemble becomes more narrowly defined, the smaller the number of configurations in the ensemble and the more highly and rapidly fluctuating the population will be. For example, if all interatomic distances in the defined assembly can only vary by 0.01% from a reference configuration, then only a very few assemblies in the system will reside within the ensemble. For a multiscale analysis to hold, and for the conventional notions of chemical kinetics as the gradual evolution of population levels to be valid, one must define an assembly type somewhat loosely. In that case the associated ensemble represents a large enough collection of $N$-atom configurations so that under appropriate conditions it changes slowly in time. Thus, vagueness in defining the assembly types becomes a part of the solution to, and not a difficulty of, the self-assembly problem. Since AMA co-evolves order parameters with the probability distribution for the atomistic configurations (notably of each of the assemblies), it is well suited for modeling self-assembly.

In the present study, we combine the above notions into a quantitative kinetic theory of self-assembly wherein population levels are the order parameters. Since the formulation is based on AMA, recalibration with each new application is unnecessary if an accurate interatomic force field and complete set of order parameters are used. Thus, the danger of over-calibration is avoided.

**Methods**

*Multiscale Analysis* The equation for stochastic dimer count ($\chi$) dynamics is obtained using an all-atom multiscale approach[24-33]. First, the Liouville equation and integration by parts imply an exact conservation law for the evolution of $\Psi$:

$$\frac{\partial \Psi}{\partial t} = -\varepsilon \frac{\partial}{\partial \chi} \int d\Gamma^* \delta(\chi - \chi^*) \rho J. \qquad (1)$$

The ansatz that the $N$-atom probability density $\rho$ takes the form $\rho(\Gamma, \chi, t)$ and thus depends on $\Gamma$ directly and, via $\chi$, indirectly is used. This dual dependence can be constructed when $\varepsilon$ is sufficiently small. The derivation presented below differs from that of other approaches because

   (a) there is no need for tedious bookkeeping to conserve the number of degrees of freedom (*6N*) despite the introduction of counter variables in $\rho$,

   (b) difficulties from integration over atomic coordinates and memory kernels in projection operator methods[41] are avoided,

(c) special ensembles constrained to fixed values of counters arise naturally[27], and

(d) use of the conservation law (1) with multiscale perturbation theory avoids the need to reconstruct $\rho$ to second order in the perturbation expansion in $\varepsilon$ to obtain corrections to them[30].

We begin the derivation by defining a new timescale structure for the system, separated in factors of $\varepsilon$ and denoted by $t_n = \varepsilon^n t$. Using the chain rule, the Liouville equation becomes

$$\sum_{n=0}^{\infty} \varepsilon^n \partial \rho / \partial t_n = \left( \mathcal{L}_0 + \varepsilon \mathcal{L}_1 \right) \rho, \tag{2}$$

$$\mathcal{L}_0 = -\sum_{i=1}^{N} \left[ \frac{\vec{p}_i}{m_i} \cdot \frac{\partial}{\partial \vec{r}_i} + \vec{F}_i \cdot \frac{\partial}{\partial \vec{p}_i} \right] \tag{3}$$

$$\mathcal{L}_1 = -J \frac{\partial}{\partial \chi}. \tag{4}$$

In this equation, $\Gamma$ derivatives in $\mathcal{L}_0$ are at constant values of $\chi$ and vice versa for $\mathcal{L}_1$. The analysis proceeds by constructing solutions of this multiscale Liouville equation as an expansion in $\varepsilon$, i.e., $\rho = \sum_{n=0}^{\infty} \rho_n \varepsilon^n$. The Liouville equation is solved to $\mathrm{O}(\varepsilon)$ and then (1) yields an equation for $\Psi$ to $\mathrm{O}(\varepsilon^2)$.

To lowest order we seek quasi-equilibrium states, i.e., for which $\rho_0$ is independent of the fastest time scale $t_0$. This is consistent with our expectation that the self-assembly of nanocomponents into composite structures takes place on timescales much longer than $10^{-14}$ seconds. In this case, the Liouville equation at $\mathrm{O}(\varepsilon^0)$ implies $\mathcal{L}_0 \rho_0 = 0$, while the $\mathrm{O}(\varepsilon)$ equation is derived to be $\left(\frac{\partial}{\partial t_0} - \mathcal{L}_0\right)\rho_1 = -\left(\frac{\partial}{\partial t_1} - \mathcal{L}_1\right)\rho_0$. Since the lowest order solution is quasi-equilibrium in character, it is determined via entropy maximization[27,28]. For an isothermal closed system, this implies

$$\rho_0 = \frac{e^{-\beta H}}{Q(\chi,\beta)} W(\chi,\underline{t}) \equiv \hat{\rho} W, \tag{5}$$

$$Q = \int d\Gamma^* \delta(\chi - \chi^*) e^{-\beta H^*} \equiv e^{-\beta F} \tag{6}$$

where $H$ is the total energy. The $\delta$-function in the expression for $Q$ arises naturally when one starts from a quantum formalism and maintains a careful count of states in the expression for the entropy. Here, $F$ is the $\chi,\beta$-dependent free energy, and from its definition, $\Psi$ is $W$ to lowest order in $\varepsilon$.

To $O(\varepsilon)$ the Liouville equation admits the formal solution $\rho_1 = e^{\mathcal{L}_0 t_0} A_1 - \int_0^{t_0} dt_0' e^{\mathcal{L}_0 (t_0 - t_0')} \left(\frac{\partial}{\partial t_1} - \mathcal{L}_1\right)\rho_0$. Here, $A_1(\Gamma,\chi,\underline{t})$, is $\rho_1$ at $t_0 = 0$ and $\underline{t} = \{t_1, t_2, \cdots\}$. With the expressions for $\mathcal{L}_1$ and $\rho_0$, and the change of variables $s = t_0' - t_0$, this implies $\rho_1 = e^{\mathcal{L}_0 t_0} A_1 - \hat{\rho} t_0 \frac{\partial W}{\partial t_1} - \hat{\rho} \int_{-t_0}^{0} ds\, e^{-\mathcal{L}_0 s} J\left(\left[\frac{\partial}{\partial \chi} - \beta \langle f \rangle\right] W\right)$ where $\langle f \rangle = -\partial F/\partial \chi$ is the thermal-average force driving $\chi$ dynamics. The Gibbs hypothesis and the self-consistency condition that $\rho_1$ must be well-behaved as $t_0 \to \infty$ together imply $\partial W/\partial t_1$ must be zero since $\langle J \rangle$, the $\hat{\rho}$-weighted average of $J$, is zero. With this,

$$\rho_1 = e^{\mathcal{L}_0 t_0} A_1 - \hat{\rho} \int_{-t_0}^{0} ds\, e^{-\mathcal{L}_0 s} J\left(\left[\frac{\partial}{\partial \chi} - \beta \langle f \rangle\right] W\right), \tag{7}$$

completing the $O(\varepsilon)$ analysis of the Liouville equation.

For arbitrary choice of $A_1$, the above analysis does not necessarily lead to a closed equation for $\Psi$, nor does $\Psi$ necessarily evolve slowly (i.e., is independent of $t_0$), violating our earlier assumption. However, if $A_1$ lies in the nullspace of $\mathcal{L}_0$, then a closed equation for the slow evolution of $\Psi$ emerges. In particular, we take the first-order initial data to be zero, i.e., the system starts in the quasi-equilibrium state $\hat{\rho}W$. With this, for later times $(t_0 > 0)$ $\Psi$ remains slowly varying suggesting self-consistency of the development. It follows that $\Psi = W$ up to $O(\varepsilon^2)$. From equations (5) and (7), the $\Psi$-conservation equation (1) implies

$$\frac{\partial \Psi}{\partial t} = \varepsilon^2 \frac{\partial}{\partial \chi}\left[\int_{-t_0}^{0} dt' \langle J e^{-\mathcal{L}_0 t'} J \rangle \left(\frac{\partial}{\partial \chi} - \beta \langle f \rangle\right)\Psi\right].$$

To reconstruct the full time dependence of $\Psi$, we express all time variables in terms of $\tau = t_2$, notably $t_0 = \tau/\varepsilon^2$. Taking the limit $\varepsilon \to 0$ of the above equation for $\partial\Psi/\partial t$ yields equations (13) and (14) of the next section. This is the Smoluchowski equation of stochastic dimerization. The diffusion coefficient depends on $\chi$, while the $\chi$-dependence of $\langle f \rangle$ drives the dimerization. At equilibrium, $\Psi$ is proportional to $e^{-\beta F}$ as expected; thus $e^{-\beta F}$ solves (13) and (14) when $\Psi$ is time-independent.

*Langevin/Smoluchowski Equivalence* We now postulate a Langevin model, given by equation (19) of the Results and Discussion section, and show that it is equivalent to the Smoluchowski equation in a Monte Carlo sense. Let $\Psi(\underline{\chi}, t)$ be the joint probability density for the set of scaled counter variables $\underline{\chi}$. We assume the set of random forces $\underline{\xi}$ changes on a more rapid timescale than $\underline{\chi}$. A time interval $\Delta t$ is postulated to exist such that the system experiences a representative sample of variations in $\underline{\xi}(t)$ during $\Delta t$. Additionally, it is assumed that the processes $\underline{\xi}$ are Markov, so that $\Psi$ can be advanced from $t$ to $t+\Delta t$ solely knowing $\Psi$ at $t$ and the probability for a transition in $\underline{\chi}$ during $\Delta t$. The evolution of $\Psi$ is determined by the statistics of all $\underline{\xi}$ timecourses between $t$ and $t+\Delta t$. Let $T[\underline{\xi},t,\Delta t]$ be the transitional probability as a functional of a scenario of $\underline{\xi}$ during the time period $t$ to $t+\Delta t$. Then, $\Psi$ evolves via

$$\Psi(\underline{\chi},t+\Delta t) = \int d^L\underline{\chi}' \mathcal{S}_{\underline{\xi}} T[\underline{\xi},t,\Delta t] \delta(\underline{\chi}-\underline{\tilde{\chi}}[\underline{\xi};t,\Delta t,\underline{\chi}'])\Psi(\underline{\chi}',t). \tag{8}$$

Here, $\underline{\tilde{\chi}}$ is the solution of (19) at time $t+\Delta t$, given the state was $\underline{\chi}'$ at $t$. The functional integral in (8) is over all scenarios of $\underline{\xi}$ between $t$ and $t+\Delta t$, and the identity $\mathcal{S}_{\underline{\xi}} T = 1$ follows from normalization of $T$.

Integration of (19) yields $\underline{\tilde{\chi}} = \underline{\chi}' + \left[\underline{\underline{\alpha}}\langle\underline{f}\rangle_{\underline{\chi}} \Delta t + \int_t^{t+\Delta t} ds\underline{\xi}(s)\right]$. Next, we insert this result for $\underline{\tilde{\chi}}$ into (8), formally expand the $\delta$-functions in the changes $\underline{\tilde{\chi}} - \underline{\chi}'$ that vanish as $\Delta t \to 0$ using $\delta(\underline{\chi}-\underline{\tilde{\chi}}) \approx \sum_{k=0}^{2} \frac{1}{k!} \frac{\partial^k \delta(\underline{\chi}-\underline{\chi}')}{\partial \underline{\chi}^k}(\underline{\chi}'-\underline{\tilde{\chi}})^k$. After this we perform the averaging implied by the $T$-weighted functional integral in (8), denoted by $\langle\ \rangle_T$. As with the reaction flux it represents, we assume the average random force vanishes, i.e., $\langle\underline{\xi}\rangle_T = \underline{0}$. If the process generating $\underline{\xi}(t)$ is stationary, then $\langle \underline{\xi}(t)\underline{\xi}(t')\rangle_T$ only depends on $t-t'$, and not on $t$ and $t'$ independently, i.e, $\langle \underline{\xi}(t)\underline{\xi}(t')\rangle_T \equiv \underline{\underline{\phi}}(t-t')$. Neglecting terms of order higher than $\Delta t$, the statistical properties of $\underline{\xi}$ and integration of (19) imply

$\langle \underline{\chi}' - \underline{\tilde{\chi}}\rangle_T = -\underline{\underline{\alpha}}\langle \underline{f}\rangle \Delta t$ and $\langle (\underline{\chi}' - \underline{\tilde{\chi}})^2\rangle_T = \int_0^{\Delta t} dt_2 \int_{-t_2'}^{\Delta t-t'} ds \underline{\underline{\phi}}(s)$. For essentially all $t'$ in $(0, \Delta t)$, the upper limit of the $s$ integral is much greater than the correlation time. Since $\underline{\underline{\phi}}(|t'|)$ is negligible except for $|t'|\ll \Delta t$, one finds $\langle (\underline{\chi}' - \underline{\tilde{\chi}})^2\rangle_T = \underline{\underline{A}}\Delta t$ where $\underline{\underline{A}} \equiv \int_{-\infty}^0 ds\underline{\underline{\phi}}(s)$.

Under the postulate that $\Psi$ changes sparingly over $\Delta t$, we find $\Psi(\underline{\chi}, t+\Delta t) \approx \Psi(\underline{\chi}, t) + \frac{\partial \Psi}{\partial t}\Delta t$. Substituting these results into (8) and neglecting terms of order higher than $\Delta t$ yields the equation $\frac{\partial \Psi}{\partial t}\Delta t = \frac{\partial}{\partial \underline{\chi}}\{\langle \underline{\chi}'-\underline{\tilde{\chi}}\rangle_T \Psi\} + \frac{1}{2}\frac{\partial^2}{\partial \underline{\chi}^2}\{\langle (\underline{\chi}'-\underline{\tilde{\chi}})^2\rangle_T \Psi\}$. The computed values for $\langle (\underline{\chi}'-\underline{\tilde{\chi}})^k\rangle_T$ for $k=1,2$ from above imply the Smoluchowski equation:

$$\frac{\partial \Psi}{\partial t} = \frac{\partial}{\partial \underline{\chi}}\left[\underline{\underline{\alpha}}\left(\frac{1}{2}\underline{\underline{\alpha}}^{-1}\underline{\underline{A}}\frac{\partial}{\partial \underline{\chi}} - \langle \underline{f}\rangle\right)\Psi\right]. \qquad (9)$$

At long times the closed, isothermal system reaches equilibrium. Hence $\Psi(\underline{\chi}, t) \underset{t\to\infty}{\sim} \frac{\exp(-\beta F)}{Z} \equiv \Psi^{eq}(\underline{\chi})$ where $Z = \int d^L\chi e^{-\beta F}$. As this must be a time-independent solution of the Smoluchowski equation, it follows that $\underline{\underline{\alpha}} = \frac{\beta \underline{\underline{A}}}{2}$.

Comparison of (13) and (14) with (9) finally implies $\beta \underline{\underline{D}} = \underline{\underline{\alpha}}$.

## Results and Discussion

*Dimerization: An Illustrative Example*    We begin by formulating the self-assembly problem for the construction of dimers and generalize in a later section. Consider dimerization of members in a collection of $M$ nanocomponents. For simplicity, assume all components contain the same number and types of atoms with the same interatomic bonding connectivity. Let $\underline{R} \left( = \{\vec{R}_1, \cdots \vec{R}_M\} \right)$ be a set of nanocomponent centers-of-mass (CMs). The notion of a dimer, even for an isolated pair of nanocomponents, is not precise. For example, how close together must the components be to constitute a dimer? Is a pair of components categorized as a dimer when a third may be colliding with them, or is this a trimer? For elongate nanocomponents, is the attachment architecture end-to-end, parallel, or perpendicular? Such ambiguities are an inherent part of the speciation problem. In fact "species" are never well-defined in a dense system, but rather are an abstraction, valid in a precise way only in extreme cases, such as a low temperature rarefied gas wherein bound dimeric quantum states can be defined. However, the impossibility of a precise structural definition of a dimer does not preclude a quantitative, operational definition of a dimeric ensemble.

Given the free energy profile of Fig. 2, it might be said that components $k$ and $k'$ form a dimer if the distance between their CMs is less than a critical value $R_c$.

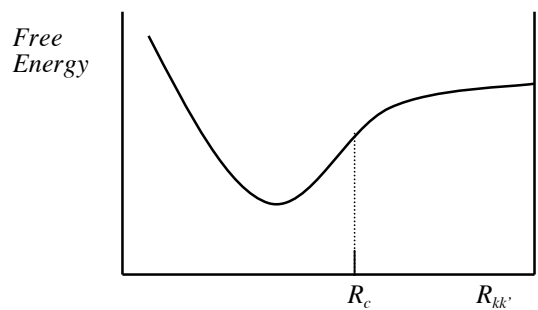

**Fig. 2**: The pseudo-pair free energy profile shows a well indicating the tendency to form a bound state with center-to-center distance at the minimum. For dimerization, the free energy has three-body character, such that a third component is repelled, or at least not attracted, to a dimer.

Such a profile represents an ensemble of atomistic configurations and not a given relative orientation or internal configuration of the two monomers. Adopting a critical CM-CM distance definition will be unsatisfactory in some cases since it does not rule out trimers and other assemblies. For components with a single small attachment patch, the three-body nature of the interaction makes trimers unlikely at low density. Thus, if two components are joined via their attachment patches, a third is sterically inhibited from attachment with them if the patch is small relative to component size. Hence, there can be simple criteria for defining a dimer when the many-body nature of the free energy-derived forces between components eliminates assemblies that are not consistent with the geometry of the components and attachment regions. The free-energy landscape may render a simple definition of a dimer inadequate. Thus, the best one can do to construct a theory of dimerization is to introduce a notion of dimeric associations, i.e., of the dynamics of an ensemble of detailed configurations with dimeric association defined in a semi-quantitative fashion. Regardless of the definition adopted, it must contain some ambiguity or there will never be an appreciable population. In fact, when one poses a classical chemical model (e.g., $2X \rightleftharpoons X_2$) it is assumed there is a dimeric association usually without being concerned with configuration detail. Thus, we develop the present approach in the same spirit.

With these cautionary notes, let a given monomer pair be classified as a dimeric association if the CMs of its two components are within the critical distance $R_c$. For this definition to suffice, it is assumed that collisions between components are inelastic. This ensures that dimers are stable for a time period that is long compared to the timescale of atomic collisions[31-33] and which naturally follows from the formalism developed.

Any approach must have a built-in self-consistency criterion to determine if a more precise definition of a dimer is required. Let a dimer population counter $X_2$ be defined via $X_2 = \sum_{k<k'} \theta(R_{kk'})$ where $R_{kk'} = |\vec{R}_k - \vec{R}_{k'}|$ and $\theta(R)$ is one when $R < R_c$ and zero otherwise. Thus, a state of dimeric association is the ensemble of all configurations of the $N$-atom system consistent with a given value of $X_2$. With these definitions, multiscale analysis can be used to derive an equation for the stochastic evolution of this ensemble.

To establish the validity of a multiscale theory of dimerization, $X_2$ must be shown to evolve on a timescale much longer than the $10^{-14}$ seconds characteristic of atomic vibrations. Newton's equations imply

$$\frac{dX_2}{dt} = \frac{1}{m_c} \sum_{k<k'} \delta(R_{kk'} - R_c) \frac{(\bar{R}_k - \bar{R}_{k'})}{R_{kk'}} \cdot (\bar{P}_k - \bar{P}_{k'}) \equiv \tilde{J}, \qquad (10)$$

where $\delta$ is the Dirac delta function, $m_c$ is the total mass of one of the identical components, and $\bar{P}_k$ is the total momentum of component $k$. Let $\varepsilon^{-2}$ be the number, $M$, of components in the system. We now show that under certain conditions a kinetic theory for $X_2$ emerges as $\varepsilon \to 0$. In the limit of large population levels, the dimer counter $X_2$ is less than $M/2$ for the single-attachment patch system at low density (i.e., $< R_c^{-3}$). Hence, it is convenient to introduce the intensive variable $\chi$ via $\chi = 2\varepsilon^2 X_2$, $0 < \chi < 1$ which represents the proportion of dimers that form in the system. Newton's equations then imply

$$\frac{d\chi}{dt} = \varepsilon J, \; J = 2\varepsilon \tilde{J}, \qquad (11)$$

for $\tilde{J}$ defined in (10). There are $O(\varepsilon^{-4})$ terms on the RHS of (10). However, there are only $O(\varepsilon^{-2})$ terms for which the $\delta$-term can be zero. The fluctuating signs of these terms due to the momentum factors $\bar{P}_k$ under near-equilibrium conditions imply that, in the net, $\tilde{J}$ is $O(\varepsilon^{-1})$. Hence, (11) implies $\chi$ evolves slowly, i.e., on a timescale of $O(\varepsilon)$. With these hypotheses and estimates, $\chi$ is therefore a slow variable. Since $J$ is $O(\varepsilon^0)$, similar arguments show that it is not slow, i.e., $dJ/dt = f$ where $f$ is $O(\varepsilon^0)$. This suggests that an equation for the stochastic dynamics of $\chi$ is of the Smoluchowski, or friction-dominated, type[29].

Unfortunately, we see that equation (10) is not closed with respect to $\chi$. To derive such an equation, a multiscale analysis is conducted beginning with the Liouville equation for the $N$-atom probability density $\rho(\Gamma,t)$, namely $\partial \rho / \partial t = \mathcal{L} \rho$ where

$$\mathcal{L} = -\sum_{i=1}^{N} \frac{\vec{p}_i}{m_i} \cdot \frac{\partial}{\partial \vec{r}_i} + F_i \cdot \frac{\partial}{\partial \vec{p}_i} \tag{12}$$

is the Liouville operator and $\Gamma = \{\vec{p}_1, \vec{r}_1, \cdots \vec{p}_N, \vec{r}_N\}$ with atomic positions and momenta denoted by $\vec{r}_i, \vec{p}_i; i=1,\ldots,N$. The reduced probability density $\Psi$ is defined via $\Psi(\chi,t) = \int d\Gamma^* \delta(\chi - \chi^*) \rho$ where $\chi^*$ is the value of the scaled dimer counter evaluated at the integration variables $\Gamma^*$.

We proceed as in our analysis of multiscale theory to other phenomena[24-33] by utilizing the multiscale ansatz that $\rho$ depends on $\Gamma$ both directly and, via $\chi$, indirectly; using the chain rule to derive an equation for $\rho$ in its $6N+N_a$ dimensional representation; expanding $\rho$ in terms of $\varepsilon$; constructing the lowest order distribution $\rho_0$ via entropy maximization; and involving the Gibbs-hypothesized equivalence of long-time and ensemble averages to ensure that the $\varepsilon$ expansion is well-behaved at long times. As a result a Smoluchowski equation for $\Psi$ emerges:

$$\frac{\partial \Psi}{\partial \tau} = \frac{\partial}{\partial \chi}\left[ D \left( \frac{\partial}{\partial \chi} - \beta \langle f \rangle \right) \Psi \right] \tag{13}$$

$$D = \int_{-\infty}^{0} ds \langle J(0) J(s) \rangle. \tag{14}$$

Here the thermal-average force $\langle f \rangle$ is the $\chi$-derivative of the free energy $F$, the bracket $\langle \cdots \rangle$ indicates an average with both canonical and $\delta(\chi - \chi^*)$ weights, $\tau$ is a scaled time variable, $J(s)$ represents the variable $J$ evolved to time $s$, and $e^{-\beta F}$ is the $\chi$-dependent partition function

$$e^{-\beta F} = \int d\Gamma^* \delta(\chi - \chi^*) e^{-\beta H^*}. \tag{15}$$

Thus, a rigorous equation for the stochastic dynamics of the dimer count, i.e. an analogue of the chemical master equation[27], emerges from the $N$-atom Liouville equation. As shown earlier in Methods, the use of the Gibbs hypothesis and all-atom multiscale analysis allows one to avoid the need for projection operators to extract $\Psi$ from the fluctuating, atomistic state. The timescale separation implied by the smallness of $\varepsilon$ yields this closed equation for $\Psi$, a feature not provided by Newton's equations directly. Since $\Psi$ depends upon $\rho$ and $\rho$ can be reconstructed from $\Psi$, our approach co-evolves the atomistic probability density with $\Psi$, i.e., is a fully coupled multiscale procedure capturing the feedback of Fig. 1.

The above results may appear to ignore the possible existence of multiple, long-lived dimeric isomers. In that case, the theory must be augmented to include these assemblies. This is critical when the various isomers evolve at distinct rates as it implies that one may have ignored slow variables other than $\chi$. Such an error would be signaled by the existence of long-time tails in $\langle J(s) J(0) \rangle$ and the anomalous (divergent) behavior of $D$. Hence, computation of $D$ provides a self-consistence check on our theory that guides the discovery of self-assembly pathways. Evaluation of the correlation function yielding *D* can be constructed using an MD package. This is not straightforward for an operator that requires memory kernels involving non-Newtonian evolution. In summary, we provide a procedure for predicting self-assembly kinetics that is calibration-free, as it is derived from the basic laws of molecular physics.

*Generalizations for Complex Assemblies* The self-consistency criterion provided by our methodology facilitates the development of a theory for self-assembly in complex systems involving aggregates of multiple nanocomponents into any of a number of alternative architectures (e.g., normal versus abnormal structures of viruses or ribosomes). If key intermediate structures, and hence assembly pathways are omitted, unphysically slow assembly kinetics will be predicted, and similarly for the creation of incorrect populations of otherwise unlikely final assemblies. The construction of counters for a $g$-type assembly (e.g., $g = 5$ for a pentamer of given architecture and range of atomic configurations) enables the development of a multiscale approach to more complex self-assembling structures. An increase in the number of components in an assembly may allow for many possible isomeric architectures. The number of different assembly intermediates depends on the geometry and interactions of the nanocomponents; the size, orientation, and positioning of attachment patches; and the possible multiple internal conformations of individual nanocomponents. This isomerization complexity could even arise for dimers as noted in the previous section. The number of possible subassembly isomers could increase exponentially with the number of monomers $M$ so that a theoretical framework for evaluating possible intermediate assemblies and pathways is of great interest. We illustrate below how criteria can be developed to ensure a more complete set of assembly intermediates, thereby leading to the discovery of key self assembly pathways. Through refinements of the algorithms for population counters, one may discriminate among a number of intermediate associations and eliminate anomalous behaviors in the associated matrix of diffusion factors.

To illustrate our pathway refinement strategy, consider a trimeric association. For a trimer, components may be arranged in a triangular versus a linear form. This can arise if each component has two sufficiently small and appropriately positioned attachment patches. Patch-to-patch interactions must have a range less than the monomer size if unbounded (i.e., supply-limited) aggregation is to occur. A possible counter for trimers constructed from identical components is $X_3 = \sum_{k>k'>k''} \theta(R_{kk'})\theta(R_{k'k''})\theta(R_{kk''})$, where $\theta$ and $R_{kk'}$ are defined as for $X_2$. For components with attachment patches positioned so as to inhibit thermally stable linear arrangements and allow only triangular ones, $X_3$ will not count linear associations since they are not present in appreciable numbers. However, care must be taken not to count dimers that are actually part of a trimer or other more complex association.

For the general aggregate of type $g$, let $X_g$ be a counter defined in analogy to that for dimers and trimers, but with additional constraints discriminating among various isomers and avoiding counting substructures of a larger structure. For example, trimer substructures of a four-cluster would not be included in $X_3$. A promising direction for developing counters for more general assemblies is as follows. Let $X_g^>$ be a counter for all aggregates with $g$ or more monomers. Then, a counter $X_g$ for $g$-mers that does not incorrectly count substructures is given by $X_g = X_g^> - X_{g+1}^>$. This addresses the problem that while it is relatively easy to construct $X_g^>$, constructing $X_g$ directly is more difficult. Within the definition of $X_g$ one can also add isomeric restrictions (e.g., a linear $g$-mer versus a globular one) via $\theta$-factors on structural variables.

The joint probability $\Psi(\underline{X},t)$ for $\underline{X} = \{X_1, X_2, \cdots, X_L\}$ in a system supporting $L$ types of assemblies is central to a stochastic theory of self-assembly. In the present framework, it is defined via $\Psi(\underline{X},t) = \int d\Gamma^* \Delta(\underline{X} - \underline{X}^*) \rho$, where $\underline{X}^* = \{X_1^*, X_2^*, ..., X_L^*\}$ is the set of counters evaluated at the all-atom state $\Gamma^*$ over which integration is taken and $\Delta$ is the $L$-fold product of delta functions. As in the derivation of (13) and (14), the Liouville equation implies

$$\frac{\partial \Psi}{\partial t} = -\underline{\nabla} \cdot \underline{J} \tag{16}$$

$$J_g = \int d\Gamma^* \Delta(\underline{X} - \underline{X}^*) \sum_{i=1}^N \frac{\vec{p}_i^*}{m_i} \cdot \frac{\partial X_g^*}{\partial \vec{r}_i^*} \rho. \tag{17}$$

One might term $J_g$ the "reaction flux" for creation of $g$-type assemblies. This result does not constitute a complete stochastic chemical kinetic theory of self-assembly unless it can be shown that $\underline{J}$ can be expressed as an explicit functional of $\Psi$ alone, and not of the full $N$-atom density. As with the dimerization problem, this requires that the set of counters $\underline{X}$ be slowly varying in time and do not couple strongly to other slow variables. Counters can also be introduced to address spatially non-uniform systems by keeping track of $\underline{X}$ for each of a number of spatial compartments or by introducing order parameter field variables[42,43].

To complete the derivation of an equation of stochastic population dynamics, the scaling arguments used for dimerization are introduced. Let the index $g$ (=1,2···,$L$) label assembly types consisting of $g$ monomers (i.e., isomeric multiplicity ignored). For a system with $M$ monomer units, $X_g$ can be at most $M/g \left( \equiv \varepsilon^2 g \right)^{-1}$. In analogy with the dimer case, one may define the scaled counter $\chi_g$ via $\chi_g = \varepsilon^2 g X_g$. Proceeding in this way, all developments set forth for dimerization follow, and we arrive at the Smoluchowski equation

$$\frac{\partial \Psi}{\partial t} = \frac{\partial}{\partial \underline{\chi}} \left[ \underline{\underline{D}} \left( \frac{\partial}{\partial \underline{\chi}} - \beta \langle \underline{f} \rangle \right) \Psi \right] \qquad (18)$$

that involves analogous diffusion coefficients $\underline{\underline{D}}$. For extremely complex systems arising when $L$ is large, say O($M$), the development must be re-evaluated.

*Langevin Simulation*      Since it is not practical to solve the Smoluchowski equation directly, Langevin equations constitute a viable way to simulate stochastic processes. Using methods of stochastic calculus[27] one may derive the

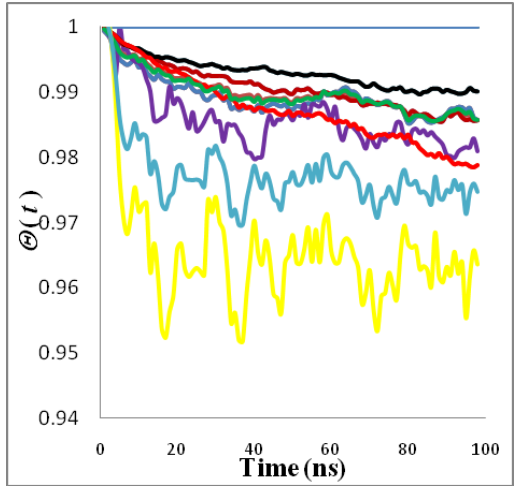

(a)

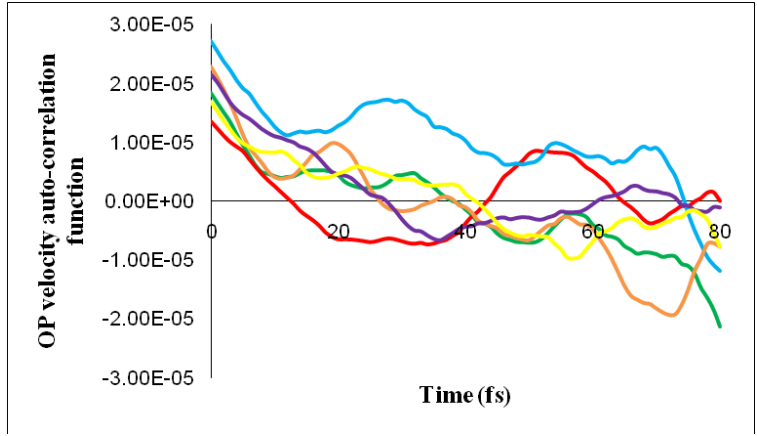

(b)

**Fig. 3**: (a) An example of order parameter time courses and (b) velocity auto-correlation functions for the OPs. Results show the absence of a long-time tail, and hence the lack of coupling to other slow variables not included in the set of OPs.

following Langevin equations that are Monte Carlo-equivalent to our Smoluchowski equation:

$$-\frac{d\underline{\chi}}{dt} + \underline{\underline{\alpha}}\langle \underline{f} \rangle + \underline{\xi} = \underline{0} \qquad (19)$$

for the set of $L$ scaled population counters $\underline{\chi} = \{\chi_1, ..., \chi_L\}$. The matrix $\underline{\underline{\alpha}}$ is associated with the diffusion factors, $\langle \underline{f} \rangle$ is the thermal-average force, and $\underline{\xi}$ is a random force. The development ultimately implies $\underline{\underline{\alpha}} = \beta \underline{\underline{D}}$, and constrains the statistics of $\underline{\xi}$ by $\underline{\underline{\alpha}} = \frac{\beta}{2} \int_{-\infty}^{0} ds \langle \underline{\xi}(0) \underline{\xi}(s) \rangle \equiv \frac{\beta}{2} \underline{\underline{A}}$. Thus there is a constraint on random forces via the integrated random force correlation functions, i.e., the classic fluctuation-dissipation relation.

The Langevin formulation can be used to conduct practical simulations of self-assembly. As $\underline{\underline{D}}$, $\langle \underline{f} \rangle$, and the properties of the random noise $\underline{\xi}$ are determined by our statistical mechanical formulas, this constitutes a calibration-free kinetic theory of self-assembly. In broad-strike, the procedure is to construct the diffusion factors and thermal-average forces for given instantaneous population levels. These are then used to perform a Langevin time step. The atomistic state must be re-established at the new population levels so that the diffusion factors and thermal-average forces can be re-computed via molecular dynamics and the cycle is repeated. If the critical distance for interaction of the nanocomponents and the density of components are small, then the calculations of the thermal-average forces and diffusion factors can be carried out on relatively small all-atom subsystems, yielding a type of domain decomposition algorithm that can be readily parallelized. Thus, the interscale feedback of Fig. 1 implies a multiscale computational algorithm for simulating the self-assembly of nanocomponents into composite structures (e.g., see Fig. 3).

*Summary and Conclusions* Smoluchowski equations for the stochastic dynamics of the population levels of intermediates along the pathway to self-assembly were derived. A necessary condition for the self-consistency of the theory is that counters for the number of each type of intermediate assembly must be sufficiently vague to ensure a statistically significant population level, but precise enough to discriminate among conformers to ensure that key pathways are not missed. A novel multiscale technique enables the derivation of Langevin-type stochastic population equations without the need to address memory kernels that arise in the projection operator method and are difficult to simulate practically. Necessary conditions for the validity of our stochastic population equations include slow evolution of the population counters relative to the timescale of atomic collisions/vibrations and the ability of counters to discriminate among key conformers. The counters must form a complete set - they cannot couple to slow variables omitted from the model. Violation of this completeness is indicated by the presence of long-time tails in the counter rate autocorrelation functions, i.e., the divergence of $\underline{\underline{D}}$. As the theory holds the promise of being parameter-free (i.e., the rate coefficients and assembly pathways are constructed via first-principles molecular physics), our methodology is predictive and quantitative. Thus it could serve as the basis of a computer-aided self-assembly process design strategy. Similar comments hold for understanding natural processes such as viral self-assembly. To fully realize the benefits of our approach, a multiscale algorithm for Langevin simulations can be implemented, as briefly described in the previous section.

**Acknowledgements.** The first author appreciates support from the United States Naval Academy under a NARC grant and the National Science Foundation under grant number DMS-098413. The second author was supported in part by the National Science Foundation (Collaborative Research in Chemistry program), National Institute of Health (NIBIB), Department Of Energy (Office of Basic Science), METAcyt, and the Indiana University College of Arts and Sciences through the Center for Cell and Virus Theory.

**Literature Cited**

1. Whitesides, G.; Boncheva, M. *Proc. Nat. Acad. Sci.* **2002,** *99*, 4769-4774.

2. Gracias, D.; Tien, J.; Breen, T.L.; Hsu, C.; Whitesides, G.M. *Science* **2000**, *289*, 1170-1172.

3. Alberts, B.; Bray, D.; Lewis, J.; Raff, M.; Roberts, K.; Watson, J.D. *Molecular Biology of the Cell*; Garland: New York, 1994.

4. Whitesides, G.; Grzybowski, B. *Science* **2002**, *295*, 2418-2421.

5. van der Schoot, P.; Zandi, R. *Phys. Bio.* **2007,** *4*, 296-304.

6. Zlotnick, A.; Stray, S.J. *Trends in Biotechnology* **2003,** *21*, 536-542.

7. Zlotnick, A., Aldrich, R.; Johnson, J.; Ceres, P.; Young, M. *Virology* **2000**, *277*, 450-456.

8. Zlotnick, A.; Johnson, J.; Wingfield, P.; Stahl, S.; Endres, D. *Biochemistry* **1999**, *38*, 14644-14652.

9. Zlotnick, A. *Journal of Molecular Recognition* **2005,** *18*, 479-490.

10. McQuarrie, D.A. *Statistical Mechanics*; Harper Collins: NY, 1976.

11. Mori, H. *Progress in Theoretical Physics* **1965,** *34,* 423-455.

12. Zwanzig, R. In *Systems Far From Equilibrium*; Garrido, L., Editor; Interscience: New York, 1980; pp. 198-221.

13. Chandrasekhar, S. *Astrophys. J.* **1943,** *97*, 255-262.

14. Bose, S.; Ortoleva, P. *J. Chem. Phys.* **1979,** *70*, 3041-3056.

15. Bose, S.; Ortoleva, P. *Physics Letters A* **1979,** *69*, 367-369.

16. Bose, S.; Bose, S.; Ortoleva, P. *J. Chem. Phys.* **1980,** *72*, 4258-4263.

17. Bose, S.; Medina-Noyola, M.; Ortoleva, P. *J. Chem. Phys.* **1981,** *75,* 1762-1771.

18. Deutch, J.M.; Oppenheim, I. *Faraday Discuss. Chem. Soc.* **1987,** *83*, 1-20.

19. Ortoleva, P. *Nonlinear Chemical Waves*; John Wiley and Sons: New York, 1992.

20. Shea, J. E.; Oppenheim, I. *J. Phys. Chem.* **1996,** *100*, 19035-19042.

21. Peters, M.H. *J. Chem. Phys.* **1998,** *110*, 528-538.

22. Peters, M.H. *J. Stat. Phys.* **1999,** *94,* 557-586.

23. Coffey, W. T.; Kalmykov, Y.; Waldron, J. *The Langevin Equation With Applications to Stochastic Problems in Physics, Chemistry and Electrical Engineering;* World Scientific Publishing: River Edge, NJ, 2004.

24. Ortoleva, P. *J. Phys. Chem.* **2005,** *109*, 21258-21266.

25. Miao, Y.; Ortoleva, P. *J. Chem. Phys.* **2006,** *125*, 044901.

26. Miao, Y.; Ortoleva, P. *J. Chem. Phys.* **2006,** *125*, 214901.

27. Pankavich, S.; Miao, Y.; Ortoleva, J.; Shreif, Z.; Ortoleva, P. *J. Chem. Phys.* **2008,** *128*, 234908.

28. Pankavich, S.; Shreif, Z.; Ortoleva, P. *Physica A* **2008,** *387*, 4053-4069.

29. Shreif, Z.; Ortoleva, P. *J. Stat. Phys.* **2008,** *130*, 669-685.

30. Shreif, Z.; Pankavich, S.; Ortoleva, P. *Phys. Rev. E* **2009,** *80*, 031703.

31. Shreif Z.; Ortoleva P. *Physica A* **2009,** *388*, 593-600.

32. Pankavich, S.; Shreif, Z,; Miao, Y.; Ortoleva, P. *J. Chem. Physics* **2009,** *130*, 194115.

33. Pankavich, S.; Ortoleva P. *J. Math. Phys.* **2010,** *51*, 063303.

34. Shea, J. E.; Oppenheim, I. *Physica A* **1997,** *247*, 417-443.

35. Morgado, W.A.M.; Oppenheim, I. *Phys. Rev. E* **1997,** *55*, 1940-1945.

36. Morgado, W.A.M.; Oppenheim, I. *Physica A* **1997,** *246*, 547-562.

37. Harries, D.; May, S.; Gelbart, W.; Ben-Shaul A. *Biophys. J.* **1998,** *75*, 159.

38. Arkhipov, A.; Freddolino, P.; Schulten K. *Structure*, **2006,** *14*, 1767-1777.

39. Chang, R.; Ayton, G.; Voth, G.; *J. Chem. Phys* **2005,** *122*, 244716.

40. Ayton, G.; Voth, G. *Journal of Structural Biology* **2007,** *157*, 570-578.

41. Zwanzig, R. *Nonequilibrium Statistical Mechanics*; Oxford University Press: United Kingdom, 2001.

42. Shreif, Z.; Adhangale, P.; Cheluvaraja, S.; Perera, R.; Kuhn, R.; Ortoleva, P. In *Lecture Notes in Computational Science and Engineering*, Springer: Netherlands, 2008; Vol 15, pp. 363-380.

43. DelleDonne, M.; Ortoleva, P. *J. Stat. Phys.* **1979,** *20*, 473-486.

**Fig. 1 Caption**: Population levels characterize the overall organization of a self-assembling system, which affects the probability of the atomistic configurations. This mediates thermal-average forces driving population dynamics, and diffusion factors mediating the rate of change and coupling among the population levels. This interscale feedback loop is central to understanding nanosystem self-assembly. The proposed multiscale approach captures this feedback by co-evolving population levels with the thermal-average forces and diffusion factors, thereby providing the essence of an algorithm for the numerical simulation of self-assembly.

**Fig. 2 Caption**: The pseudo-pair free energy profile shows a well indicating the tendency to form a bound state with center-to-center distance at the minimum. For dimerization, the free energy has three-body character, such that a third component is repelled, or at least not attracted, to a dimer.

**Fig. 3 Caption**: (a) An example of order parameter sample paths and (b) velocity auto-correlation functions for the OPs. Results show the absence of long-time tails, and hence the lack of coupling to other slow variables not included in the set of OPs

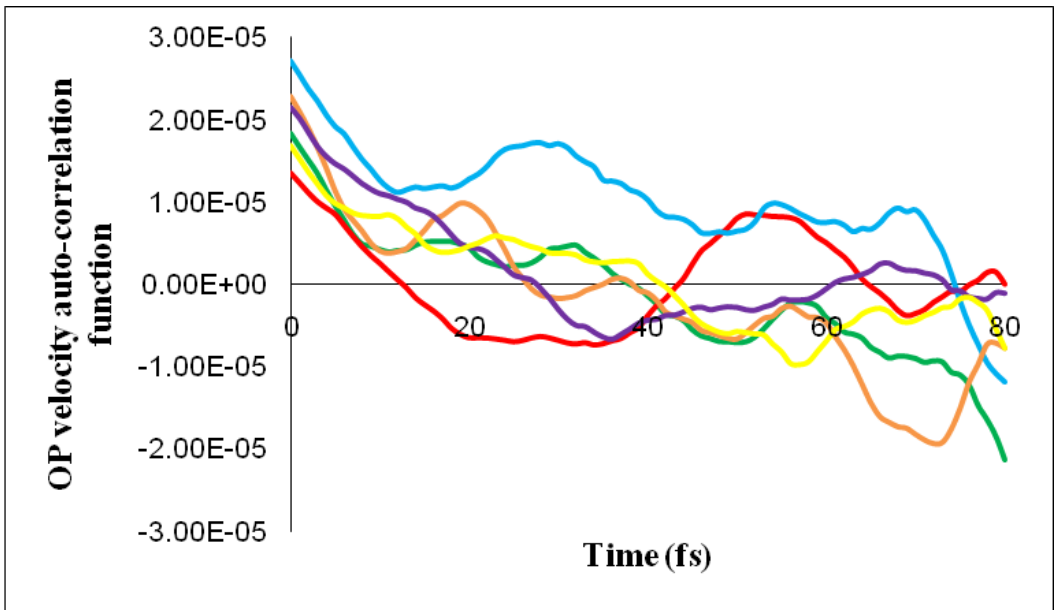

TOC graphic